\documentstyle[12pt,axodraw,epsf,epsfig]{article}
\advance\textheight by \topskip
\newcommand{\sla}{\kern -5.4pt /}
\newcommand{\Dir}{\kern -6.4pt\Big{/}}
\newcommand{\Dirin}{\kern -10.4pt\Big{/}\kern 4.4pt}
\newcommand{\DDir}{\kern -7.6pt\Big{/}}
\newcommand{\DGir}{\kern -6.0pt\Big{/}}

\newcommand{\ra}{\rightarrow}
\newcommand{\be}{\begin{equation}}
\newcommand{\ee}{\end{equation}}
\newcommand{\bea}{\begin{eqnarray}}
\newcommand{\eea}{\end{eqnarray}}
\newcommand{\beanon}{\begin{eqnarray*}}
\newcommand{\eeanon}{\end{eqnarray*}}
\newcommand{\ba}{\begin{array}}
\newcommand{\ea}{\end{array}}
\newcommand{\bi}{\begin{itemize}}
\newcommand{\ei}{\end{itemize}}
\newcommand{\ben}{\begin{enumerate}}
\newcommand{\een}{\end{enumerate}}
\newcommand{\bc}{\begin{center}}
\newcommand{\ec}{\end{center}}

\newcommand{\bfig}{\begin{center}\begin{picture}}
\newcommand{\efig}[1]{\end{picture}\\{\small #1}\end{center}}
\newcommand{\flin}[2]{\ArrowLine(#1)(#2)}
\newcommand{\wlin}[2]{\DashLine(#1)(#2){2.5}}

\newcommand{\glin}[3]{\Photon(#1)(#2){2}{#3}}

\newcommand{\sof}{\SetOffset}
\newcommand{\bmip}[2]{\begin{minipage}[t]{#1pt}\bfig(#1,#2)}
\newcommand{\emip}[1]{\efig{#1}\end{minipage}}

\newcommand{\ZP}[1]{{\it Z.\ Phys.\ }{\bf #1}}

\newcommand{\PR}[1]{{\it Phys.\ Rev.\ }{\bf #1}}

\begin{document}
\tolerance=100000
\thispagestyle{empty}
\setcounter{page}{0}

\begin{flushright}
{\large DFTT 55/96}\\ 
{\large  PSI--PR--96--23}\\
{\rm September 1996\hspace*{.5 truecm}}\\ 
hep-ph/9609468
\end{flushright}

\vspace*{\fill}

\bc     
{\Large \bf CC10 at ${\cal O}(\alpha_s )$: QCD corrections to
$e^+ e^- \to \mu^- \bar \nu_\mu~u~\bar d$ at LEP2 and the
Next Linear Collider. 
\footnote{ Work supported in part by Ministero 
dell' Universit\`a e della Ricerca Scientifica.\\[2 mm]
e-mail: maina@to.infn.it, pittau@psw218.psi.ch, pizzio@to.infn.it}}\\[2.cm]
{\large Ezio Maina$^{a}$, Roberto Pittau$^{b}$ and Marco Pizzio$^{a}$}\\[.3 cm]
{\it $^{a}$Dip. di Fisica Teorica, Universit\`a di Torino
     and INFN, Sezione di Torino,}\\
{\it v. Giuria 1, 10125 Torino, Italy.}\\[5 mm]
{\it$^{b}$Paul Scherrer Institute}\\
{\it CH-5232 Villigen-PSI, Switzerland}

\ec

\vspace*{\fill}

\begin{abstract}
{\normalsize
\noindent
QCD one-loop corrections to the semileptonic process
$e^+ e^- \to \mu^- \bar \nu_\mu~u~\bar d$ are computed.
We compare the exact calculation with a ``naive''approach to strong 
radiative corrections which has been widely used in the literature
and discuss the phenomenological relevance of QCD
contributions for LEP2 and NLC physics.
}
\end{abstract}

\vspace*{\fill}
\newpage
\section*{Introduction}
The measurement of $W$--pair production at LEP2 will provide two additional
pieces of information to our understanding of the Standard Model (SM)
\cite{lep2}.
First, it will improve the determination of the $W$ mass; second, it will probe
the structure of  triple gauge--boson couplings (TGC's).
The mass of the $W$ boson in the SM is tightly constrained.
In fact an indirect determination of $M_W$ can be obtained from a global fit of
all electroweak data. The fit gives
\be
M_W = 80.359 \pm 0.051^{+0.013}_{-0.024} {\rm GeV}
\ee 
where the central value correspond to $M_H = 300$ GeV and
the second error reflects the change of $M_W$ when the Higgs mass
is varied between 60 and 1000 GeV. 
A more precise determination of $M_W$ will provide a stringent test of the SM.
A disagreement between the value of $M_W$ derived from the global fit and the
value extracted from direct measurement would represent a major failure of the
SM. Alternatively, an improvement in the value of the $W$ mass can 
significantly tighten present bounds on the Higgs mass.
\par
Two methods have been singled out as the most promising \cite{Wmass}.
The first one is based on the rapid increase of the total cross section at
threshold. The second method relies on the direct
reconstruction of the mass from the hadronic decay products of the $W$ using
the decay channels
\bea
W^+ W^- &\ra& q \overline{q} \ell \nu \label{semilep}\\
W^+ W^- &\ra& q_1 \bar q_1 q_2 \bar q_2 \label{had}
\eea
where $\ell = e, \mu$. 
\par
Preliminary studies indicate that the direct measurement
will provide a more precise determination of the $W$ mass than the
threshold method for which a smaller number of events will be available.
Combining all decay channels, an accuracy of about 35
MeV is expected from direct reconstruction, while the ultimate precision
attainable at threshold is estimated to be about 100 MeV.

In $WW$ production triple gauge--boson couplings, which play a central role in
non--abelian gauge theories, appear already at tree level and can be studied in
much more detail than at lower energies \cite{WTGC}.
At LEP2 TGC's will be mainly probed using angular distributions of the $W$'s
and of their decay products. At the higher energies available at the Next Linear
Collider (NLC) it will be possible in addition 
to study TGC's using the energy dependence
of the total production cross section, since non--standard TGC's in general lead
to a cross section which increases with energy and therefore eventually violates
unitarity.

On--shell $W$--pair production is described at tree level by a set of three 
diagrams, labeled (e) and (f) in fig. 1. At the lowest level of sophistication, 
one can attach to them the on--shell decay of the 
$W$'s. In this case all one--loop electroweak and strong radiative
corrections are known \cite{EWcorr1, EWcorr2, QCDcorr}.
This approach is gauge invariant, but all effects of the finite
width of the $W$ and all correlations between the two decays are neglected.
In order to take these features into account one can consider the set of
diagrams mentioned above with off--shell $W$'s, which decay to four fermions.
In the literature the corresponding amplitude is called CC3. CC3 however is not
gauge invariant. A gauge independent description of $WW$ production
requires in the unitary gauge ten (twenty) diagrams for the semileptonic channel
(\ref{semilep}) when $\ell =  \mu \ (e)$ and eleven diagrams for the hadronic
channel (\ref{had}).
These amplitudes are known as CC10 (CC20) and CC11 respectively. Numerically the
total cross sections obtained from CC3 and CC10 or CC11 at LEP2 and NLC
energies differ by a few per mil. Larger discrepancies, of the order of several
per cent, are found when comparing CC3 with CC20.

In order to extract the desired information from $WW$ production data,
theoretical prediction with uncertainties smaller than those which are foreseen
in the experiments are necessary. In particular this means that radiative
corrections have to be under control.

In this letter we present the complete calculation of QCD corrections to CC10.
In most studies they have been included ``naively'' with the substitution 
$\Gamma_W \ra \Gamma_W (1 +2\alpha_s/\pi/3)$ and multiplying the hadronic
branching ratio by $(1 +\alpha_s/\pi)$. This prescription is exact for CC3 when
fully inclusive quantities are computed. However it can only be taken as 
an order of magnitude estimate even for CC3 in the presence of cuts on the jet
directions, as discussed for the hadronic channel in \cite{MP}
\footnote{ The impact of QCD corrections on the angular distribution of the
decay products of a $W$ and their application to on--shell $W$--pair production
is discussed in ref.\cite{lampe}}.
It is well known that differential distributions can be more sensitive to 
higher order corrections than total cross--sections in which virtual and real
contributions tend to cancel to a large degree.
It is therefore necessary to include higher order QCD effects into the
predictions for $WW$ production and decay in a way which allows to impose
realistic cuts on the structure of the observed events.
\vfil
\bfig(160,245)
\sof(-140,140)
\flin{45,5}{65,25}\flin{65,25}{45,45}
\Text(42,5)[rt]{$e^-$}
\Text(42,45)[rb]{$e^+$}
\glin{65,25}{85,25}{4}
\Text(75,30)[b]{$Z, \gamma$}
\sof(-155,140)
\flin{145,45}{125,65}\flin{125,65}{145,85}
\wlin{113.5,38.5}{125,65}
\flin{120,5}{100,25}\flin{100,25}{120,45}
\Text(149,85)[lb]{$\mu^-$}
\Text(149,45)[lt]{$\bar \nu_\mu$}
\Text(124,45)[lb]{$u$}
\Text(124,5)[lt]{$\bar d$}
\Text(90,0)[t]{(a)}
\sof(-10,140)
\flin{45,5}{65,25}\flin{65,25}{45,45}
\Text(42,5)[rt]{$e^-$}
\Text(42,45)[rb]{$e^+$}
\glin{65,25}{85,25}{4}
\Text(75,30)[b]{$Z$}
\sof(-25,140)
\flin{145,45}{125,65}\flin{125,65}{145,85}
\wlin{113.5,38.5}{125,65}
\flin{120,5}{100,25}\flin{100,25}{120,45}
\Text(149,85)[lb]{$u$}
\Text(149,45)[lt]{$\bar d$}
\Text(124,45)[lt]{$\mu^-$}
\Text(124,5)[lt]{$\bar \nu_\mu$}
\Text(90,0)[t]{(b)}
\sof(120,180)
\flin{45,5}{65,25}\flin{65,25}{45,45}
\Text(42,5)[rt]{$e^-$}
\Text(42,45)[rb]{$e^+$}
\glin{65,25}{85,25}{4}
\Text(75,30)[b]{$Z,\gamma$}
\sof(105,180)
\flin{120,5}{100,25}\flin{100,25}{120,45}
\wlin{113.5,11.5}{125,-15}
\flin{145,-35}{125,-15}\flin{125,-15}{145,5}
\Text(124,45)[lb]{$\mu^-$}
\Text(124,5)[lb]{$\bar \nu_\mu$}
\Text(149,5)[lb]{$u$}
\Text(149,-35)[lt]{$\bar d$}
\Text(90,-40)[t]{(c)}
\sof(-140,60)
\flin{45,5}{65,25}\flin{65,25}{45,45}
\Text(42,5)[rt]{$e^-$}
\Text(42,45)[rb]{$e^+$}
\glin{65,25}{85,25}{4}
\Text(75,30)[b]{$Z,\gamma$}
\sof(-155,60)
\flin{120,5}{100,25}\flin{100,25}{120,45}
\wlin{113.5,11.5}{125,-15}
\flin{145,-35}{125,-15}\flin{125,-15}{145,5}
\Text(124,45)[lb]{$u$}
\Text(124,5)[lb]{$\bar d$}
\Text(149,5)[lb]{$\mu^-$}
\Text(149,-35)[lt]{$\bar \nu_\mu$}
\Text(90,-40)[t]{(d)}
\sof(-10,40)
\flin{45,5}{65,25}\flin{65,25}{45,45}
\Text(42,5)[rt]{$e^-$}
\Text(42,45)[rb]{$e^+$}
\glin{65,25}{85,25}{4}
\Text(75,30)[b]{$Z,\gamma$}
\wlin{85,25}{97,45}
\wlin{85,25}{97,5}
\flin{117,32}{97,45}\flin{97,45}{117,65}
\flin{117,-15}{97,5}\flin{97,5}{117,18}
\Text(121,65)[lb]{$\mu^-$}
\Text(121,32)[l]{$\bar \nu_\mu$}
\Text(121,18)[l]{$u$}
\Text(121,-15)[lt]{$\bar d$}
\Text(75,-20)[t]{(e)}
\sof(120,60)
\flin{65,-15}{65,25}\flin{65,25}{45,45}
\flin{45,-35}{65,-15}
\Text(42,-35)[rt]{$e^-$}
\Text(42,45)[rb]{$e^+$}
\wlin{65,25}{95.5,25}
\wlin{65,-15}{95.5,-15}
\flin{110.5,13}{95.5,25}\flin{95.5,25}{110.5,45}
\flin{110.5,-35}{95.5,-15}\flin{95.5,-15}{110.5,-3}
\Text(114.5,45)[lb]{$u$}
\Text(114.5,13)[l]{$\bar d$}
\Text(114.5,-3)[l]{$\mu^-$}
\Text(114.5,-35)[lt]{$\bar \nu_\mu$}
\Text(75,-40)[t]{(f)}
\efig{Figure 1: { Tree level diagrams for 
$e^+ e^- \to \mu^- \bar \nu_\mu~u~\bar d$. The dashed lines are $W$'s.}}
\clearpage

\bfig(160,220)
\sof(-100,170)
\flin{45,0}{25,20}\flin{25,20}{45,40}
\wlin{0,20}{25,20}
\GOval(25,20)(6,6)(0){1}
\Text(25,20)[]{1}
\Text(50,20)[l]{$=$}
\flin{110,0}{90,20}\flin{90,20}{110,40}
\wlin{70,20}{90,20}
\Gluon(105,5)(105,35){-2}{4}
\sof(-100,100)
\flin{45,0}{25,20}\flin{25,20}{45,40}
\glin{0,20}{25,20}{6}
\wlin{25,20}{50,20}
\GOval(25,20)(6,6)(0){1}
\Text(25,20)[]{2}
\Text(63,20)[l]{$=$}
\sof(-100,95)
\glin{85,25}{100,25}{4} \flin{125,0}{100,25} \flin{100,25}{125,50}
\wlin{119,44}{125,63}
\Gluon(110,15)(110,35){-2}{3}
\Text(135,25)[l]{$+$}
\sof(-30,95)
\glin{85,25}{100,25}{4} \flin{125,0}{100,25} \flin{100,25}{125,50}
\wlin{115,40}{125,63}
\Gluon(120,5)(120,45){-2}{5}
\Text(135,25)[l]{$+$}
\sof(40,95)
\glin{85,25}{100,25}{4} \flin{125,0}{100,25} \flin{100,25}{125,50}
\wlin{115,40}{125,63}
\GlueArc(112.5,37.5)(10,-135,45){2}{4}
\Text(135,25)[l]{$+$}
\sof(110,95)
\glin{85,25}{100,25}{4} \flin{125,0}{100,25} \flin{100,25}{125,50}
\wlin{120,45}{125,63}
\GlueArc(110,35)(10,-135,45){2}{4}
\sof(-100,30)
\flin{45,0}{25,20}\flin{25,20}{45,40}
\glin{0,20}{25,20}{6}
\wlin{25,20}{50,20}
\GOval(25,20)(6,6)(0){1}
\Text(25,20)[]{3}
\Text(63,20)[l]{$=$}
\sof(-100,25)
\glin{85,25}{100,25}{4} \flin{125,0}{100,25} \flin{100,25}{125,50}
\wlin{119,6}{125,-13}
\Gluon(110,15)(110,35){-2}{3}
\Text(135,25)[l]{$+$}
\sof(-30,25)
\glin{85,25}{100,25}{4} \flin{125,0}{100,25} \flin{100,25}{125,50}
\wlin{115,10}{125,-13}
\Gluon(120,5)(120,45){-2}{5}
\Text(135,25)[l]{$+$}
\sof(40,25)
\glin{85,25}{100,25}{4} \flin{125,0}{100,25} \flin{100,25}{125,50}
\wlin{115,10}{125,-13}
\GlueArc(112.5,12.5)(10,-45,135){2}{4}
\Text(135,25)[l]{$+$}
\sof(110,25)
\glin{85,25}{100,25}{4} \flin{125,0}{100,25} \flin{100,25}{125,50}
\wlin{120,5}{125,-13}
\GlueArc(110,15)(10,-45,135){2}{4}
\efig{Figure 2: { Basic combinations of loop diagrams. All virtual QCD
corrections to electroweak four fermion processes can be computed using
these three sets. The quark
wave functions corrections are not included because they vanish, in
the massless limit, using dimensional regularization.}} 

\section*{Calculation}
One-loop virtual QCD corrections to 
$e^+ e^- \to \mu^- \bar \nu_\mu~u~\bar d$
are obtained by dressing all diagrams in fig. 1 with gluon loops.  
Defining suitable combinations of diagrams as in fig. 2
one can organize all contributions in a very modular way, as
shown in fig. 3. 
Note that  the first and the fourth contributions in fig. 3
can be obtained from each other by multiplying by -1 and
interchanging momenta and helicities of $u$ and $\bar d$.

The calculation has been performed using standard 
Passarino-Veltman techniques \cite{PV} and dimensional regularization for
ultraviolet, collinear and soft divergences. With the help
of the Symbolic Manipulation program FORM \cite{FORM}, all
tensorial integrals have been reduced to linear combinations of scalar 
loop functions. One needs to calculate one four-point, one two-point and 
three three-point basic functions with different input momenta, 
so that twenty independent scalar loop functions contribute 
to the cross section.

Having classified loop corrections as in fig. 2, one
easily convinces oneself that exactly the same ingredients appear
in the computation of ${\cal O}(\alpha_s)$ virtual corrections to any
electroweak four-fermion process. In fact, due to the color structure,
gluons connecting different spinor lines in the final state start
contributing at ${\cal O} (\alpha_s^2)$. Therefore, all that is required
in order to extend our results to the calculation 
of QCD corrections for all possible electroweak four-fermion
final states is the computation of the real gluon emission amplitudes. 
We plan to pursue this program in the near future.

\bfig(160,240)
\sof(-140,135)
\flin{45,5}{65,25}\flin{65,25}{45,45}
\Text(42,5)[rt]{$e^-$}
\Text(42,45)[rb]{$e^+$}
\glin{65,25}{85,25}{4}
\Text(75,32)[b]{$Z, \gamma$}
\sof(-155,135)
\flin{145,45}{125,65}\flin{125,65}{145,85}
\wlin{98,25}{125,65}
\flin{120,5}{100,25}\flin{100,25}{120,45}
\Text(149,85)[lb]{$\mu^-$}
\Text(149,45)[lt]{$\bar \nu_\mu$}
\Text(124,45)[lb]{$u$}
\Text(124,5)[lt]{$\bar d$}
\GOval(100,25)(6,6)(0){1}
\Text(100,25)[]{2}
\sof(-10,135)
\flin{45,5}{65,25}\flin{65,25}{45,45}
\Text(42,5)[rt]{$e^-$}
\Text(42,45)[rb]{$e^+$}
\glin{65,25}{85,25}{4}
\Text(75,30)[b]{$Z$}
\sof(-25,135)
\flin{145,45}{125,65}\flin{125,65}{145,85}
\wlin{113.5,38.5}{125,65}
\flin{120,5}{100,25}\flin{100,25}{120,45}
\Text(149,85)[lb]{$u$}
\Text(149,45)[lt]{$\bar d$}
\Text(124,45)[lt]{$\mu^-$}
\Text(124,5)[lt]{$\bar \nu_\mu$}
\GOval(125,65)(6,6)(0){1}
\Text(125,65)[]{1}
\sof(120,175)
\flin{45,5}{65,25}\flin{65,25}{45,45}
\Text(42,5)[rt]{$e^-$}
\Text(42,45)[rb]{$e^+$}
\glin{65,25}{85,25}{4}
\Text(75,30)[b]{$Z,\gamma$}
\sof(105,175)
\flin{120,5}{100,25}\flin{100,25}{120,45}
\wlin{113.5,11.5}{125,-15}
\flin{145,-35}{125,-15}\flin{125,-15}{145,5}
\Text(124,45)[lb]{$\mu^-$}
\Text(124,5)[lb]{$\bar \nu_\mu$}
\Text(149,5)[lb]{$u$}
\Text(149,-35)[lt]{$\bar d$}
\GOval(125,-15)(6,6)(0){1}
\Text(125,-15)[]{1}
\sof(-140,55)
\flin{45,5}{65,25}\flin{65,25}{45,45}
\Text(42,5)[rt]{$e^-$}
\Text(42,45)[rb]{$e^+$}
\glin{65,25}{85,25}{4}
\Text(75,32)[b]{$Z,\gamma$}
\sof(-155,55)
\flin{120,5}{100,25}\flin{100,25}{120,45}
\wlin{97,25}{125,-15}
\flin{145,-35}{125,-15}\flin{125,-15}{145,5}
\Text(124,45)[lb]{$u$}
\Text(124,5)[lb]{$\bar d$}
\Text(149,5)[lb]{$\mu^-$}
\Text(149,-35)[lt]{$\bar \nu_\mu$}
\GOval(100,25)(6,6)(0){1}
\Text(100,25)[]{3}
\sof(-10,35)
\flin{45,5}{65,25}\flin{65,25}{45,45}
\Text(42,5)[rt]{$e^-$}
\Text(42,45)[rb]{$e^+$}
\glin{65,25}{85,25}{4}
\Text(75,30)[b]{$Z,\gamma$}
\wlin{85,25}{97,45}
\wlin{85,25}{97,5}
\flin{117,32}{97,45}\flin{97,45}{117,65}
\flin{117,-15}{97,5}\flin{97,5}{117,18}
\Text(121,65)[lb]{$\mu^-$}
\Text(121,32)[l]{$\bar \nu_\mu$}
\Text(121,18)[l]{$u$}
\Text(121,-15)[lt]{$\bar d$}
\GOval(97,5)(6,6)(0){1}
\Text(97,5)[]{1}
\sof(120,55)
\flin{65,-15}{65,25}\flin{65,25}{45,45}
\flin{45,-35}{65,-15}
\Text(42,-35)[rt]{$e^-$}
\Text(42,45)[rb]{$e^+$}
\wlin{65,25}{95.5,25}
\wlin{65,-15}{95.5,-15}
\flin{110.5,13}{95.5,25}\flin{95.5,25}{110.5,45}
\flin{110.5,-35}{95.5,-15}\flin{95.5,-15}{110.5,-3}
\Text(114.5,45)[lb]{$u$}
\Text(114.5,13)[l]{$\bar d$}
\Text(114.5,-3)[l]{$\mu^-$}
\Text(114.5,-35)[lt]{$\bar \nu_\mu$}
\GOval(95.5,25)(6,6)(0){1}
\Text(95.5,25)[]{1}
\efig{Figure 3: One loop gluonic corrections to
$e^+ e^- \to \mu^- \bar \nu_\mu~u~\bar d$.}

The real emission  contribution for $e^+ e^- \to \mu^- \bar \nu_\mu~u~\bar d$
can be obtained attaching a gluon to the quark line of the diagrams shown in
fig. 1 in all possible positions. This results in twenty--four diagrams.
The required matrix elements have been
computed using the formalism presented in ref. \cite{method} with the help of a
set of routines (PHACT) \cite{phact} which generate the building blocks of the
helicity amplitudes  semi-automatically. 

If we write the full NLO cross section in the form 
\be\label{prima}
\sigma_{NLO}= \int_5 d\sigma^R + \int_4 d\sigma^V ,
\ee 
where we assume that all ultraviolet divergencies have been canceled by
renormalization, the real (R) and virtual (V) contributions are still
separately singular in four dimensions because of soft and collinear
singularities, while the sum is finite.

In order to be able to integrate separately the real and virtual part 
one has to explicitly cancel all singular contributions
in each term in a consistent way.  
To this aim we have used the subtraction method: the full expression 
(\ref{prima}) is rewritten in the form 
\be \label{seconda}  
\sigma_{NLO} = \int_5 [ d\sigma^R - d\sigma^S
] +  \int_4  d\sigma^V  +  \int_5  d\sigma^S.
\ee 
The subtraction term $ d\sigma^S$ must  have the same
pointwise singular behaviour as the exact real emission matrix element in order
to cancel soft and collinear divergencies. It must also be possible
to integrate $ d\sigma^S$ analytically in {\it d} dimensions over
the one parton subspace which generates the singularities.
The result of this integration is then summed to the virtual contribution
producing a finite remainder that can be treated in four dimensions. 
Benefits of this  method are twofold. First, an exact result is obtained and
no approximation needs to be taken; second, all singular terms are canceled 
under the integration sign and not at the end of the calculation,
leading to better numerical accuracies. This is especially relevant for the
present case, since we are aiming for high precision results, with errors
of the order of a per mil.
We have found it particularly convenient to 
implement the recently proposed dipole formul\ae\,\cite{catani}. These
are a set of completely general factorization expressions 
which interpolate smoothly between the soft eikonal factors and the 
collinear Altarelli--Parisi kernels in a Lorentz covariant way, hence
avoiding any problem of double counting in the region in which partons are both 
soft and collinear. 

All integrations have been carried out using the Montecarlo routine VEGAS 
\cite{vegas}.

An important ingredient for accurate predictions of $W$--pair production is the
effect of electromagnetic radiation. In the absence of a calculation of all 
${\cal O}(\alpha )$ corrections to four--fermion processes, these effects can
only be included partially. 
In contrast with LEP1 physics a gauge invariant separation of initial and final
state radiation is not possible. On the other hand, 
the leading logarithmic part of initial state
radiation is gauge invariant and can be included using structure
functions. The non--logarithmic terms however are unknown. 
In order to assess the influence of QCD corrections, we are
interested in a comparison with the results obtained in the Workshop on Physics
at LEP2. Hence we have decided
to employ as much as possible the parameters adopted in the 
``tuned comparisons'' \cite{WMC} 
which provide the most extensive collection of results. Therefore we have used
the $\beta$ prescription in the structure functions, where
$\beta = \ln(s/m^2) - 1$. In the same spirit we have not included Coulomb
corrections to CC3, which are known to have a sizable effect, particularly
at threshold. They could however be introduced with minimal effort.

\begin{figure}
\bc
\begin{tabular}{|c|c|}
\hline
 \rule[-7 pt]{0 pt}{24 pt}
parameter & value \\
\hline \hline
 \rule[-7 pt]{0 pt}{24 pt}
$M_Z$ & 91.1888 GeV \\ \hline
 \rule[-7 pt]{0 pt}{24 pt}
$\Gamma_Z$ & 2.4974 GeV \\ \hline
 \rule[-7 pt]{0 pt}{24 pt}
$M_W$ & 80.23 GeV \\ \hline
 \rule[-7 pt]{0 pt}{24 pt}
$\Gamma_W$ & $3 G_F M_W^3/(\sqrt{8}\pi )$ \\ \hline
 \rule[-7 pt]{0 pt}{24 pt}
$\alpha^{-1}=\alpha^{-1}(2 M_W)$ & 128.07 \\ \hline
 \rule[-7 pt]{0 pt}{24 pt}
$G_F$ & 1.16639$\,\cdot\, 10^{-5}$ Gev$^{-2}$ \\ \hline
 \rule[-7 pt]{0 pt}{24 pt}
$\sin^2\theta_W$ & $\pi\alpha (2 M_W)/(\sqrt{2} G_F M_W^2)$\\ \hline
 \rule[-7 pt]{0 pt}{24 pt}
$\alpha_s$ & .117 (.123) \\ \hline
 \rule[-7 pt]{0 pt}{24 pt}
$V_{CKM}$ & 1 \\ \hline
\end{tabular}
\ec
\bc
Table I: Input parameters.
\end{center}
\end{figure}

\section*{Results}
In this section we present a number of cross sections and of distributions
for $e^+ e^- \to \mu^- \bar \nu_\mu~q_1~\overline {q_2}$. In all cases we sum
over the two possibilities $(q_1, q_2) = ( u, d)$ and $(q_1, q_2) = ( c, s)$.
The input parameters used in our calculation are given in table I. At LEP2
energies we have used $\alpha_s = .117$ as in \cite{MP}, while at the NLC we
have adopted $\alpha_s = .123$ in order to conform to the choice made for the
Joint ECFA/DESY Study: Physics and Detectors for a Linear Collider.
Initial state radiation is included in all results.

\begin{figure}
\bc
\begin{tabular}{|c|c|c|c|} \hline 

$\sqrt{s}$ & Born & Exact QCD & ``naive'' QCD \\ \hline\hline
 \rule[-7 pt]{0 pt}{24 pt}
$161$ GeV &.24962 $\pm$ .00002 &.24760 $\pm$ .00002 &.24790 $\pm$ .00002\\ 
\hline
 \rule[-7 pt]{0 pt}{24 pt}
$175$ GeV &.96006 $\pm$ .00007 &.94519 $\pm$ .00007 &.94613 $\pm$ .00007\\ 
\hline
 \rule[-7 pt]{0 pt}{24 pt}
$190$ GeV &1.184003 $\pm$ .00009 &1.16681 $\pm$ .00009 &1.16766 $\pm$ .00008\\ 
\hline
 \rule[-7 pt]{0 pt}{24 pt}
$500$ GeV &.46970 $\pm$ .00006 &.47109 $\pm$ .00007&.46131 $\pm$ .00006 \\ 
\hline
\end{tabular}
\ec
\bc
Table II: Cross sections in pb with canonical cuts (see text).
\ec
\end{figure}

Previous studies \cite{WMC} have shown that the differences between the total 
cross sections
obtained from CC10 and those obtained with CC3 are at the per mil level.
Much larger effects have been found in observables like the average shift of the
mass reconstructed from the decay products from
the true $W$ mass. If $s_-$ and $s_+$ are the invariant masses of the 
$\mu^- \bar \nu_\mu$ pair  and of the hadronic system, respectively,
the standard definition is:
\be
\langle \Delta M\rangle = \frac{1}{\sigma} \int 
\left( \frac{\sqrt{s_+} +\sqrt{s_-} - 2\,M_W}{2E_b}\right) d\sigma.
\ee
This quantity vanishes in the zero width approximation and provides
a useful estimate of
the influence of various physical processes on the relationship between
the measured value of the $W$ mass and its actual value.

For LEP2 we have adopted the so called ADLO/TH set of cuts:
\bi
\item $E_\mu\ > \ 1$ GeV; \hspace{1 cm} 
               $10^\circ \ < \ \theta_\mu \ < \ 170^\circ $
\item $ \theta_{\mu j} \ < \ 5^\circ $
\ei
Furthermore:
\bi
\item the energy of a jet must be greater than 3 GeV;
\item two jets are resolved if their invariant mass is larger than 5 GeV;
\item jets can be detected in the whole solid angle.
\ei

For the NLC we have adopted the NLC/TH set of cuts which differ from the ADLO/TH
set in that a minimum angle of $5^\circ$ is required between a jet and either
beam and that two jets are resolved if their invariant mass is larger
than 10 GeV. Both set of cuts will also be referred to as ``canonical'' in the
following.

The assumption that each final state particle corresponds to a jet must be
abandoned when going from LO to NLO calculation. 
Starting from the final state partons, it is necessary to define jets using
an infrared safe procedure. Only in this case the cancellation of infrared and
collinear singularities between virtual and real corrections can take place and
meaningful results can be obtained. In the present case it is natural to
define jets using the ADLO(NLC)/TH cuts. Following
this prescriptions we have  merged  into one jet those parton pairs whose
invariant mass was smaller than 5(10) GeV. Furthermore partons with energy
below 3 GeV have been merged using the JADE algorithm. 
Having identified jets, we have checked whether they passed 
the canonical cuts.
All events with two or three observed jets have been retained in our plots
and cross sections.

In fig. 4 we present the normalized distribution of the angle between the muon
and the closest jet at the three energies at which data will be taken at LEP2,
$\sqrt{s} = 161,\ 175$ and 190 GeV. 
The corresponding cross sections are given
in table II together with the results at Born level and those obtained
with ``naive'' QCD (nQCD) corrections.
The events in fig. 4 pass all ADLO/TH cut with 
the exception of the minimum $\theta_{\mu j}$. Notice that imposing the
latter cut corresponds to discarding only the leftmost bin in the plot.
According to the ADLO/TH prescription, jets can be measured over the full solid
angle. Therefore, apart from the consequences of the increase in the $W$ width,
the only possible effects of QCD radiation in the semileptonic channel are 
related to the increased probability of a jet to be close to
the charged lepton. With increasing collider energy, the two $W$'s tend to fly
apart with larger relative momentum and therefore the probability of a jet to
overlap with the lepton decreases. This behaviour is clearly visible in fig. 4
and is confirmed by fig. 6a which
shows that at $\sqrt{s} = 500$ GeV jets are typically well separated in angle
from the charged lepton and that the differences between the exact distribution
and the one with ``naive'' corrections is confined to very large angles.
From fig. 4 and 6a it is also apparent that, as expected, exact QCD predicts
smaller minimum angle between jets and the charged lepton.
nQCD results for this observable become closer to the exact NLO 
distribution as the energy increases.

Table II shows that QCD corrected cross sections for 
$e^+ e^- \to \mu^- \bar \nu_\mu~q_1~\overline {q_2}$ are between 1\%, at 
$\sqrt{s} = 161$ GeV, to 2\% smaller, at $\sqrt{s} = 190$ GeV, than 
Born prediction at LEP2 with ADLO/TH cuts.
In the absence of cuts the exact NLO result for the total cross section and the
``naive'' implementation of QCD corrections agree to better than $10^{-4}$ and
are indistinguishable within statistical integration errors for all energies
studied in table II. This is expected
since the two inclusive cross sections differ by corrections of order 
$\alpha_s/\pi$ to terms of relative order $10^{-3}$.
After cuts the two sets of results at LEP2 differ by about $10^{-3}$,
while statistical
errors are smaller than $10^{-4}$. Therefore, not surprisingly, there is clear
evidence that nQCD fails even for cross sections when phase space for
final particles is limited by cuts. However, with LEP2 canonical cuts the
difference is much smaller than the projected experimental accuracy.
Because of the additional cut on the minimum angle between jets and initial
state leptons, measurements at the NLC are far less inclusive than at LEP2.
The effect is further amplified by the high--energy behaviour of $W$--pairs
which tend to be produced at smaller angles that near threshold.
Therefore at $\sqrt{s} = 500$ GeV, with NLC/TH cuts, the exact NLO cross section
is about 2\% larger than what is obtained with nQCD, a difference which
is much larger than the expected experimental uncertainties.
It is amusing to note that in this case the exact NLO result is very close,
about .5\% larger, to the tree level cross section than to the nQCD 
prediction, which is approximately 1.5\% smaller than the Born result.
The decrease in cross section due to the larger width of the $W$ is compensated
by the extra radiation which leads to a larger number of events with two or more
visible jets.

The mass shift $\langle \Delta M\rangle$ at $\sqrt{s} = 175$ GeV, with
canonical cuts, is found to be
\be
\langle \Delta M\rangle_{NLO} = -0.6383\cdot 10^{-2} \pm 0.0002\cdot 10^{-2}
\ee
which happens to be in excellent agreement with the nQCD result 
$\langle \Delta M\rangle_{nQCD} = -0.6381\cdot 10^{-2} \pm 0.0002\cdot 10^{-2}$.
This is to be compared with the tree level result of 
$\langle \Delta M\rangle_{Born} = -0.6219\cdot 10^{-2} \pm 0.0002\cdot 10^{-2}$.
Therefore the large effect, about 2.5\%,
of QCD corrections on $\langle \Delta M\rangle$,
which was suggested by the ``naive'' approach \cite{WMC}, 
is confirmed by our calculation and placed on a solid footing.

Fig. 5 shows the distribution of the minimum angle between any jet
and either beam at $\sqrt{s} = 175$ GeV. Even though the ADLO/TH set does
not include a requirement on the $\theta_j$ angle, such a cut is included in the
experimental studies for the hadronic channel, and is part of the
NLC/TH set. We have separated the contribution of two--jet and three--jet final
states making it possible to estimate the effect of different angular cuts
on the cross section.

In fig. 6 two distribution at $\sqrt{s} = 500$ GeV are presented. Fig. 6a
shows the distribution of the angle between the muon and the closest jet,
while fig. 6b shows the distribution of the minimum angle between any jet
and either beam. 
Fig. 6a makes it clear that essentially all events would pass any reasonable
isolation cut for the $\mu^-$.
In fig. 6b we separate again the contribution of two--jet and three--jet final
states so that the effects of non--canonical cuts can be judged from the plot.

\section*{Conclusions}
We have described the complete calculation of QCD radiative corrections to 
to the semileptonic process $e^+ e^- \to \mu^- \bar \nu_\mu~u~\bar d$ which are
essential in order to obtain theoretical predictions for $W$--pair production
with per mil accuracy. The amplitudes we have derived are completely
differential, and realistic cuts can be imposed on the parton level
structure of the observed events.
It has been shown that the ``naive'' implementation of QCD corrections
fails in the presence of cuts on the direction of final state jets. The error is
typically at the per mil level at LEP2.
However at $\sqrt{s} = 500$ GeV, with NLC/TH cuts, the discrepancy is 
about 2\%, much larger than the expected experimental precision.
QCD corrections substantially increase, by more than 2\%, the average shift 
between the $W$ mass measured from the decay products and the actual value
of $M_W$.

\vfill\eject

\newpage
\thispagestyle{empty}
\centerline{
\epsfig{figure=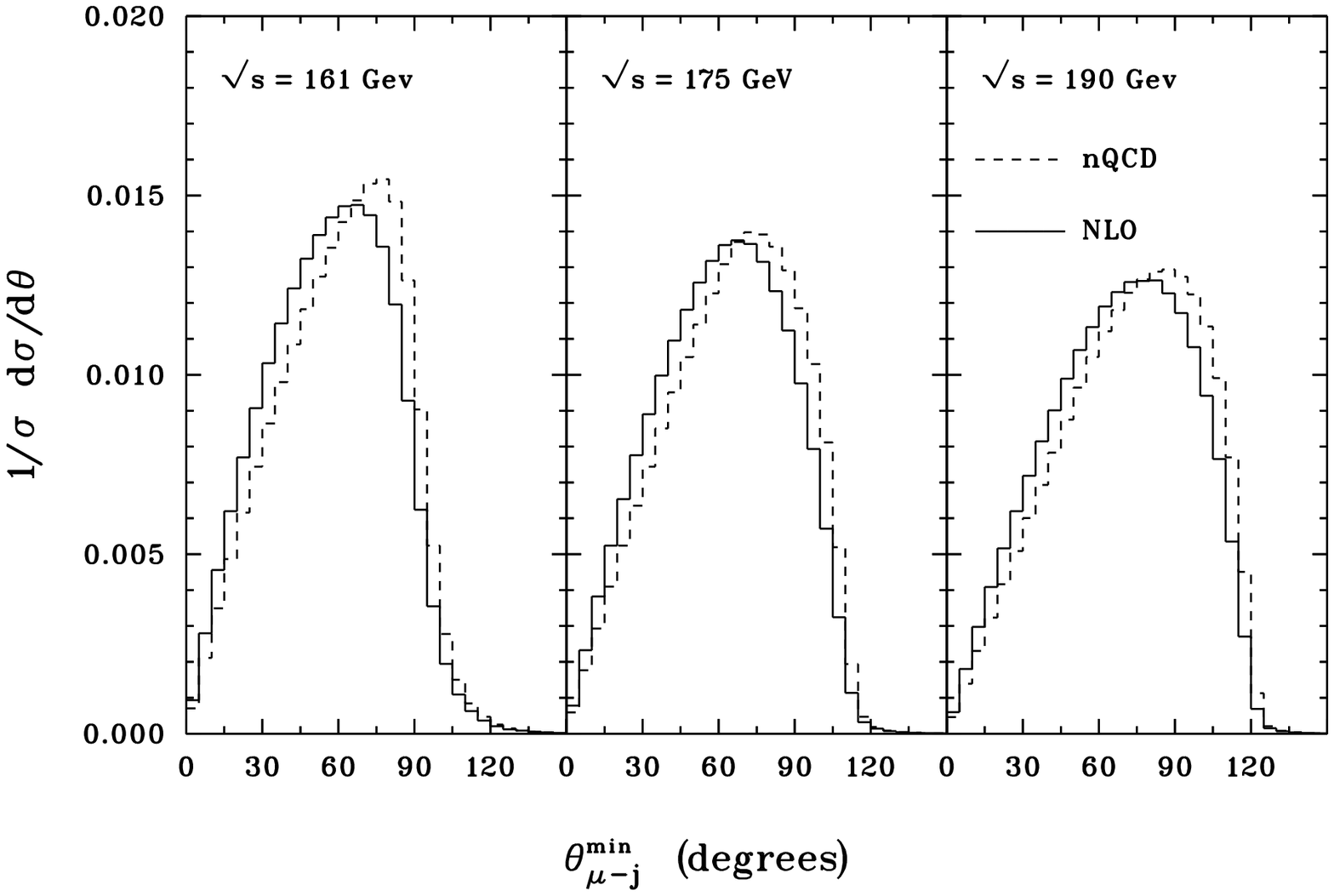,width=16cm}
}
\vspace{-6.7 cm}
\noindent
Fig. 4: Distribution of the angular separation of the $\mu^-$ from
the closest jet at $\sqrt{s} = 161, \ 175, \ 190$ GeV. All ADLO/TH cuts with the
exception of that on $\theta^{\rm min}_{\mu - {\rm jet}}$ are applied.
The continuous histogram is the exact NLO result while the dashed histogram 
refers to nQCD.

\newpage
\thispagestyle{empty}
\centerline{
\epsfig{figure=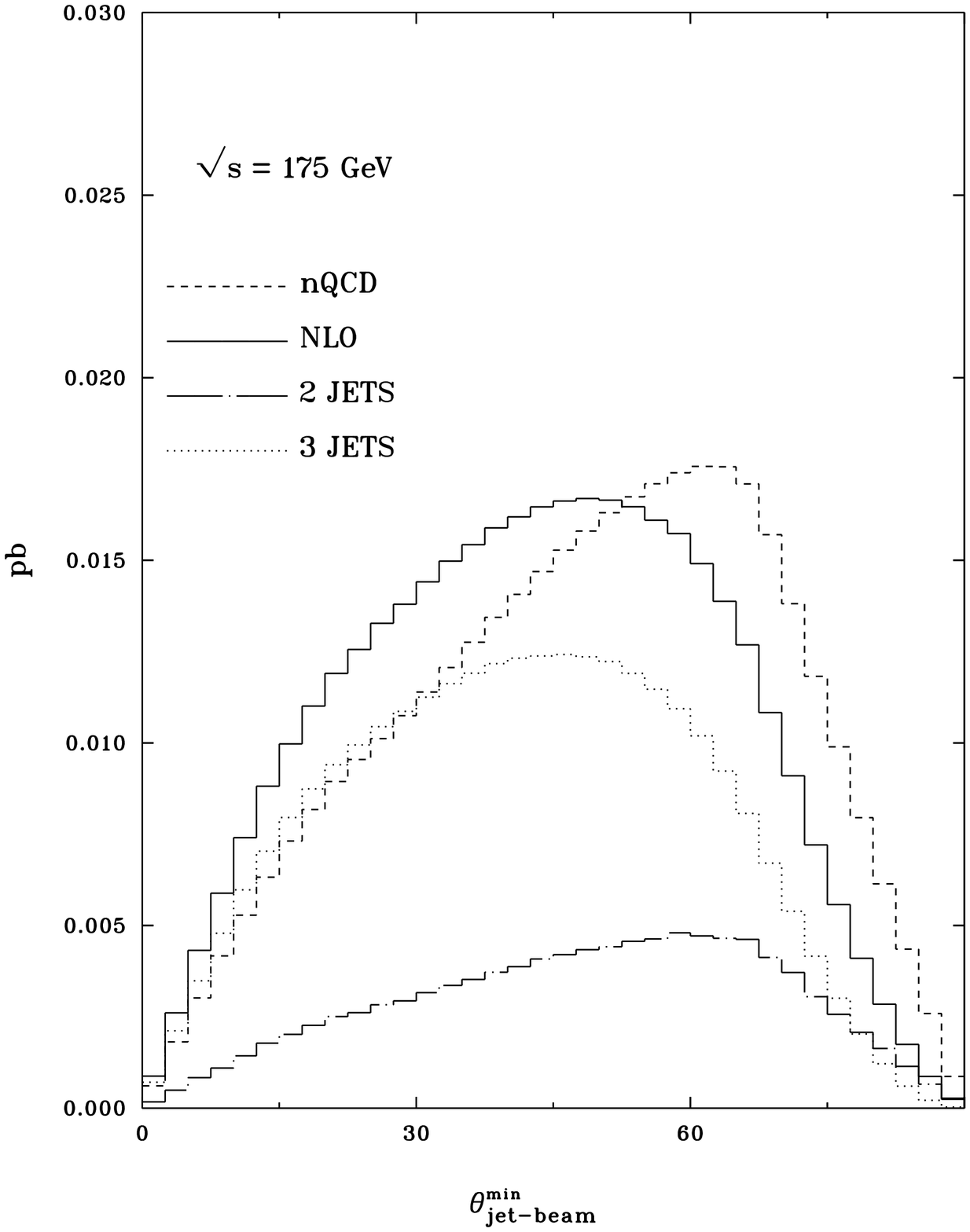,width=14 cm}
}
\vspace*{1 cm}
\noindent
Fig. 5: Distribution of the minimum angular separation between any
 jet and either beam at $\sqrt{s} = 175$ GeV with canonical cuts.
The continuous, dotted and dot--dashed histograms are exact NLO results
while the dashed histogram refers to nQCD.

\newpage
\thispagestyle{empty}
\centerline{
\epsfig{figure=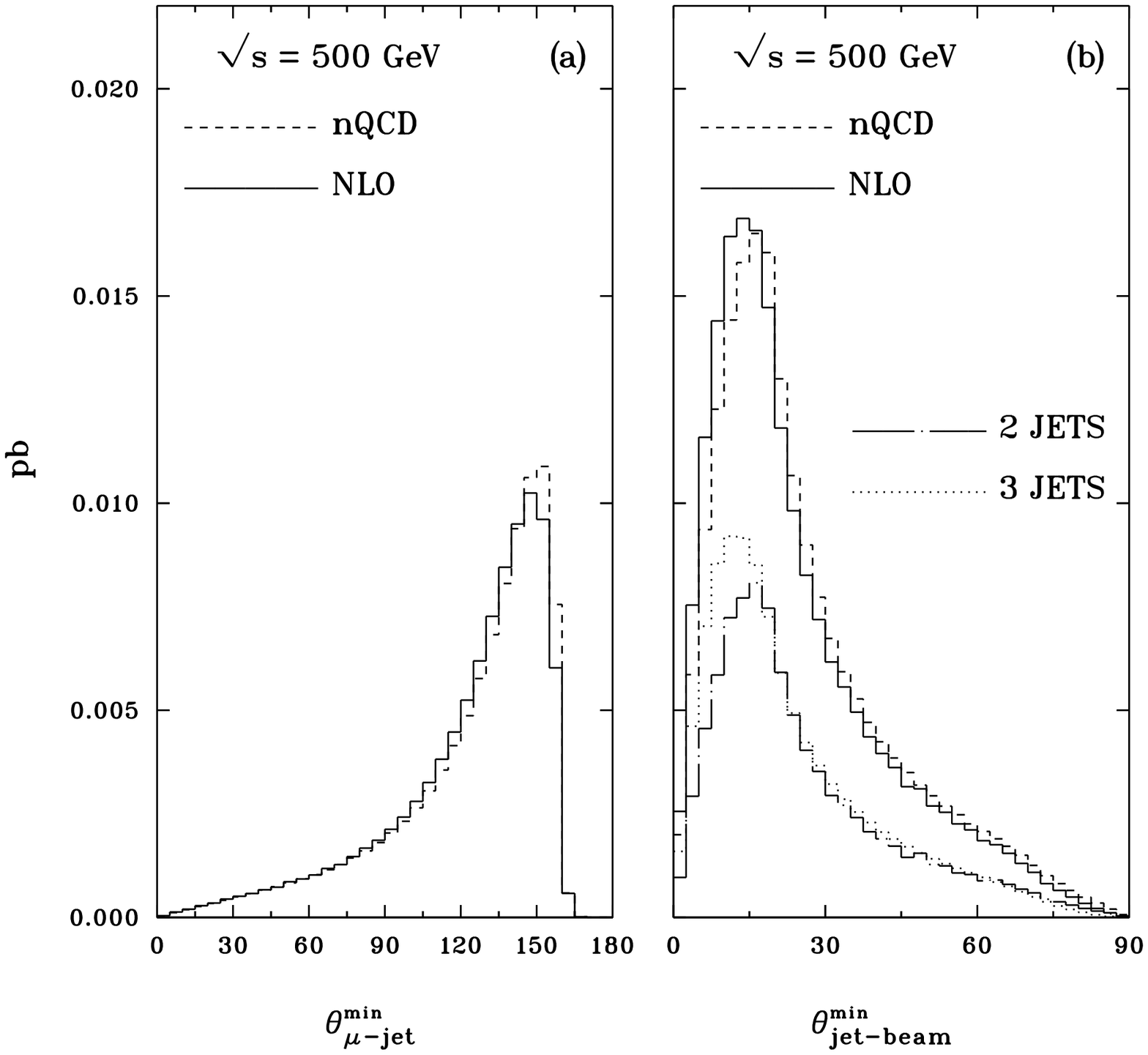,width=15 cm}
}
\vspace*{-3.5 cm}
\noindent
Fig. 6: Distribution of the angular separation of the $\mu^-$ from
the closest jet (a) and of the minimum angular separation between any
jet and either beam (b) at $\sqrt{s} = 500$ GeV.
The continuous, dotted and dot--dashed histograms are exact NLO results
while the dashed histogram refers to nQCD.


\begin{thebibliography}{1}

\bibitem{lep2} The most complete review of $W$--pair production at LEP2 can be
found in the Proceedings of the Workshop on Physics at LEP2, G.~Altarelli,
T.~Sj\"ostrand and F. Zwirner eds., Cern~96-01.

\bibitem{Wmass} A.~Ballestrero {\it et al.}, Determination of the mass of the
$W$ boson, in ref.\cite{lep2}, Vol. 1, p. 141.

\bibitem{WTGC} Z.~Ajaltouni {\it et al.}, Triple gauge boson coupling, 
in ref.\cite{lep2}, Vol. 1, p. 525.

\bibitem{EWcorr1} M.~B\"ohm {\it et al.}, {\it Nucl. Phys.} {\bf B 304} (1988)
463;\\
W.~Beenakker, K.~Ko{\l}odziej and T.~Sack, Phys.~Lett. {\bf B258} (1991) 469;\\
W.~Beenakker, F.A.~Berends and T.~Sack, {\it Nucl. Phys.} {\bf B 367} (1991)
287.

\bibitem{EWcorr2} J.~Fleischer, F.~Jegerlehner and M.~Zra{\l}ek, \ZP{C42} (1989)
409;\\
K.~Ko{\l}odziej and M.~Zra{\l}ek, \PR{D43} (1991) 3619;\\
J.~Fleischer, F.~Jegerlehner and K.~Ko{\l}odziej, \PR{D47} (1993) 830.

\bibitem{QCDcorr} D.~Albert, W.J.~Marciano, D.~Wyler and Z.~Parsa, 
{\it Nucl. Phys.} {\bf B 166} (1980) 460.

\bibitem{MP} E.~Maina and M.~Pizzio, Phys.~Lett. {\bf B369} (1996) 341.

\bibitem{lampe} K.J.~Abraham and B.~Lampe, MPI--PhT/96--14, hep-ph/9603270.

\bibitem{PV} G.~Passarino and M.~Veltman, {\it Nucl. Phys.} {\bf B 160}
(1979) 151.

\bibitem{FORM} J.A.M.~Vermaseren, ``FORM'', Computer Algebra Nederland,
Amsterdam 1991.

\bibitem{catani}
S.~Catani and M.H.~Seymour, preprint CERN-TH/96-29 (hep-ph/9605323).

\bibitem{method}
A.~Ballestrero and E.~Maina, Phys.~Lett. {\bf B350} (1995) 225.

\bibitem{phact} A.~Ballestrero, in preparation.

\bibitem{vegas} G.P.~Lepage, Jour. Comp. Phys.  {\bf 27} (1978) 192.

\bibitem{WMC} E.~Accomando {\it et al.}, Event generators for WW physics,
in ref.\cite{lep2}, Vol. 2, p. 3.

\end{thebibliography}
\end{document}